# Hashigo: A Next-Generation Sketch Interactive System for Japanese Kanji

Paul Taele and Tracy Hammond

Sketch Recognition Laboratory
Department of Computer Science
Texas A&M University
College Station, TX 77843-3112
ptaele@cse.tamu.edu, (979) 845-7143
hammond@cse.tamu.edu, (979) 862-4284

**Abstract**

Language students can increase their effectiveness in learning written Japanese by mastering the visual structure and written technique of Japanese kanji. Yet, existing kanji handwriting recognition systems do not assess the written technique sufficiently enough to discourage students from developing bad learning habits. In this paper, we describe our work on Hashigo, a kanji sketch interactive system which achieves human instructor-level critique and feedback on both the visual structure and written technique of students' sketched kanji. This type of automated critique and feedback allows students to target and correct specific deficiencies in their sketches that, if left untreated, are detrimental to effective long-term kanji learning.

## Introduction

One of the major difficulties faced by students of Japanese as a Second Language (JSL) is mastering the three scripts in written Japanese. This holds especially true for students with strong fluency in English, since they must deal with written scripts that have no resemblance to the Latin alphabet utilized in languages such as English. The kanji (literally, "Chinese characters") script particularly poses the most difficult problem to JSL students, due to its non-phonetic properties.

In addition, JSL students must experience a steep learning curve and make a long-term investment in order to achieve sufficient fluency in kanji comprehension. This is due to the script's vast character set that numbers in the thousands, its complicated visual structure involving strokes that can number to at least thirty, a high similarity between characters in the set, creating "shape collisions" during memorization, and to the divergent writing styles of native Japanese users (Yang 1998). The fact that JSL students must have a working knowledge of no less than two thousand kanji before they can effectively communicate with native Japanese writers (Lin 2007) further emphasizes the importance of kanji instruction in the JSL curriculum.

In order to help students overcome these difficulties, JSL programs traditionally introduce various written techniques (e.g., stroke order, number, and direction) to ease the kanji learning process (Banno *et al.* 1999) and to provide a systematic way for JSL students to memorize kanji more efficiently. In addition, written technique usage is greatly stressed early on in kanji learning to discourage the development of bad learning habits (McNaughton & Ying 1999). Acquiring correct written technique habits early in a student's kanji learning greatly eases the memorization process; this is especially important when students encounter the much more complex kanji in advanced-level Japanese language courses. The added benefit to improved kanji learning is that students will have additional time to focus on other important aspects of the Japanese language (e.g., conversation practice).

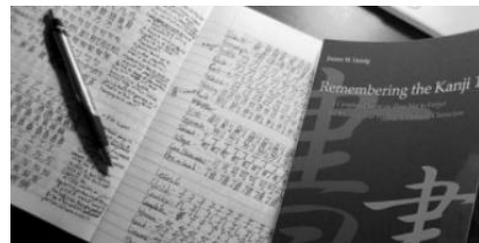

**Figure 1.** Typical paper-based assignment for practicing kanji.

Although expertise in written technique is just as important as visual structure during the course of kanji learning, JSL programs still currently employ paper-based workbooks and assignments as the primary tools to critique student performance. Language instructors are limited to gauging student visual structure performance on these traditional paper-based tools, since only the final result of the written kanji can be evaluated. One workaround is for instructors to physically monitor students writing kanji; but, this solution is increasingly time-consuming on the instructors' part as the number of kanji introduced and students in the classroom intensifies. Another method is for instructors to require students to label and enumerate the strokes in their written technique. Not only is this solution unnatural in the writing process, but it also increases the workload on the students that could have been better spent acquiring new kanji. Lastly, instructors could leave the written technique learning literally in the hands of the students, but this solution is equivalent to abandoning direct written technique instruction entirely.

With improving hardware and decreasing costs of Tablet PCs, computer-assisted instruction (CAI) systems with digital sketching capabilities are becoming viable options to augment traditional kanji teaching in JSL programs. An appropriate CAI system for kanji teaching would provide active learning (i.e., learning by doing instead of viewing or listening) that fosters interactive involvement, with the goal of expanding beyond the limitations of paper-based tools such as textbooks and workbooks (van Dam *et al.* 2005). Learning kanji through writing is important because it allows students to "improve the aesthetic appearance of their writing and acquire a 'natural feel' for the flow of the kanji that cannot be achieved simply by remembering them" (Heisig 2007). Therefore, it is highly beneficial for a CAI system to incorporate sketch interaction capabilities, as well as to provide automated feedback and assessments on the visual structure and written technique of students' written kanji.

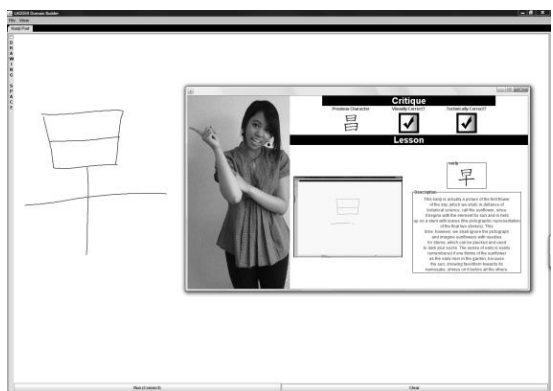
**Figure 2.** Our kanji-based sketch interactive system.

In this paper, we discuss our CAI system, Hashigo: a fully functional sketch interactive system for augmenting the teaching of kanji. With this system, students can obtain the same kind of valuable feedback on their visual structure and written technique that instructors provide. Hashigo automates the task of assessing written technique, discouraging bad learning habits in JSL students early on and allowing instructors more time to focus on other important aspects of the language.

## Related Work

A key component in our sketch interactive system is recognizing how and which kanji a student is writing; a process which is itself a pattern recognition problem. Therefore, we focus the discussion in this section on both prevalent online kanji recognition approaches and related kanji-based CAI systems.

### Online Kanji Recognition Approaches

Numerous research works concentrate solely on online recognition approaches of kanji, and approaches developed in this field have been incorporated in kanji-based CAI systems (Nakagawa 1999). Two of the most common approaches for online kanji recognition are vision-based and gesture-based, both of which differ greatly in how they handle recognition of handwritten kanji.

**Vision-Based Approaches.** Online kanji recognition systems which utilize vision-based approaches focus on classifying handwritten kanji based on features extracted solely from the kanji's visual structure. Since the means of writing the kanji are not taken into account, vision-based approaches are free from dealing with written technique properties associated with what the user wrote (Liu *et al.* 2004). This is a desirable property for systems whose core users are native or expert kanji writers, since accuracy rates do not suffer when, for example, users write kanji with an alternative stroke order, or with a non-standard number of strokes.

Disregarding written technique during the recognition process omits the very information used to provide novice JSL students with the feedback necessary to improve their kanji writing; vision-based approaches for kanji recognition, therefore, are ill-suited for kanji-based CAI systems. One consequence to JSL students of omitting this information is that, even in cases when they write visually-correct kanji with incorrect technique, the system will misinform the student that the kanji was written completely correct. Another consequence that would displease instructors is that, even if JSL students do not complete the writing of the kanji or, instead, write them sloppily, the system may still inform students that their written kanji is correct. This type of feedback from these systems may impede JSL students' progress in acquiring correct written technique.

**Gesture-Based Approaches.** Unlike vision-based approaches, gesture-based approaches take into account how users write kanji. These approaches typically classify handwritten kanji by: First, sampling the points; then, extracting the features or segmenting the lines; next, codifying the stroke directly (e.g., indexing) or assigning them probabilistically (e.g., hidden Markov models); after that, determining how those features or lines interrelate; and finally, determining their hierarchical structure (i.e., its composition of simpler characters, if there are any) (Liu *et al.* 2004).

Even though gesture-based approaches retain information regarding how users write kanji, they cannot distinguish an error due to the user drawing the wrong character from an error due to improper stroke error. This generalization is very frustrating from a human-computer interaction stand-point. Imagine if a student is asked to draw the kanji for 'water.' If the student drew the correct kanji, but with the wrong stroke order, it may not be obvious to the student where the source of the error lies, and learning may become a guessing game for the student.

## Kanji-Based CAI Systems

Given the complexity of teaching kanji to JSL students with native or primary English fluency, instructors welcome kanji-based CAI systems that could aid students in efficiently absorbing many kanji. Most current kanji-based CAI systems, including a representative sampling listed in (Lin *et al.* 2007) and (Fujita *et al.* 2002), do not allow for sketching in the learning process – nor do they use artificial intelligence, as is used in our method, to teach kanji – rather, they use standard audio or visual modalities, asking the user to repeat or identify characters on the screen. Since kanji is a purely written component of the Japanese language, the major limitation posed by these systems is the failure to incorporate sketch interactivity in the teaching process.

While other kanji-based CA systems do exist that utilize gesture- or vision-based techniques to recognize students' sketches (Nakagawa *et al.* 1999) (Lin *et al.* 2004), none of these systems are capable of providing distinct feedback for visual technique and stroke order. Thus, these systems are not effective for, nor were they designed for, teaching novice JSL students. Our Hashigo system differs from currently utilized online kanji recognition approaches and existing kanji-based CAI systems in that it aims to emulate a JSL instructor's capability in critiquing students' kanji for visual structure and written technique, and it uses sketch recognition and artificial intelligence to solve a task trivial to humans but currently difficult for computers.

Previous research work in (Taele & Hammond 2008a) and (Taele & Hammond 2008c) demonstrated the viability of employing geometric-based sketch recognition methods for visual structure recognition of Chinese characters. In addition, research work in (Taele & Hammond 2008b) showed that written technique recognition is achievable for the related domain of Mandarin Phonetic Symbols I. The distinction between this work and the work above is not only one of language (Chinese versus Japanese); previous work described only the feasibility of such an approach in a simply prototype, whereas this system describes a complete teaching tool to be integrated into the Japanese Language Lab at our university.

## Implementation

In order to provide the visual structure and written technique critique in kanji teaching that human instructors already do well, we expand upon existing gesture-based kanji recognition methods and incorporate geometric-based sketch recognition methods. Students' written kanji are first checked for visual structure correctness, Then, temporal and spatial information from the written kanji's raw data are post-processed for written technique correctness.

### Resources

The key assumption we make with kanji in general is that they can be approximated entirely with lines. Using this assumption, we make use of two tools to aid in the construction of our kanji-based CAI system. The first tool is a corner-finding algorithm developed by Sezgin, which allows us to segment students' written kanji into their primitive line components, using velocity and curvature data from the digital stylus (Sezgin *et al.* 2001). Our key assumption holds even when the written kanji contains slight curves, since our set threshold values allow the algorithm to approximate these curves as lines, except at extreme curvature.

The importance of fragmenting written kanji into line components comes into play with our usage of the second tool, the LADDER sketching language (Hammond & Davis 2005). In our CAI system, kanji that we wish to recognize are defined through geometric constraints, using the language constructs found in LADDER. The result is that written input will be classified as a specific kanji if the input's visual properties fulfill the geometric constraints written for that kanji. Furthermore, the LADDER sketching language allows us to explicitly assign labels to the individual lines and their containing endpoints and midpoints. This labeling capability allows us to critique the written technique of students' kanji by comparing temporal information of extracted line components to their corresponding assigned labels. We elaborate on this later.

### Pre-processing: Shape Descriptions

Visual structure critique using the LADDER sketching language requires us to construct shape descriptions for each kanji we wish to recognize. We utilize three types of specifications that make up shape descriptions in LADDER, which are: components, constraints, and aliases. Visual structures and partial shape descriptions for representative kanji can be found in Figures 3 and 4.

The first specification we make use of is components, which consist of a combination of predefined and user-defined shapes that serve as the building blocks of the kanji we wish to recognize. Predefined shapes in LADDER consist of primitive shapes such as lines, arcs, curves, and ellipses, while user-defined shapes consist of shapes in the domain that the user creates. For the domain we're working with, lines are the only predefined shape used, while partial and complete kanji encompass user-defined shapes. The more complex kanji shape descriptions contain simpler kanji, as shown for the "ancient" kanji in Figure 3d.

The next specification we utilize is constraints, which explicitly defines the behaviors related to the components and the relationships between them. The format style we use for constraints is divided into line orientations, endpoint ordering, and spatial relationships. Since we approximate kanji entirely as lines, we constrain the lines as either sloped, anti-diagonal, or the negation of one of the two. Also, each line in LADDER has endpoints and midpoints assigned *p1*, *p2*, and *center*, respectively; so, we explicitly define the locations of endpoints relative to each other. Once we constrain the lines, we finally define the interrelationships between the line components.

The last specification we employ is aliases, which allows us to assign alternate labels to existing component names. One of the benefits of using aliases is their ability to provide more intuitive names to either lines or points in those lines. Therefore, we can label a particular line either by its physical feature or by its stroke enumeration. Labeling by its stroke enumeration is also vital to written technique recognition in our system, and shall be elaborated further in the next section.

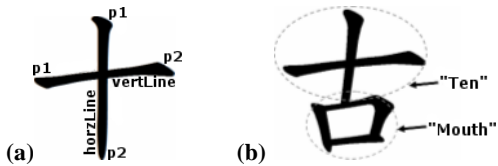

| (c) Partial Shape Description for 十 | | |
|---|---|---|
| **components:** | | |
| Line | horzLine | |
| Line | vertLine | |
| **constraints:** | | |
| Horizontal | horzLine | |
| Vertical | vertLine | |
| LeftOf | horzLine.p1 | horzLine.p2 |
| RightOf | vertLine.p1 | vertLine.p2 |
| SameSize | horzLine | vertLine |
| SameX | horzLine.center | vertLine.center |
| SameY | horzLine.center | vertLine.center |
| **aliases:** | | |
| Point | leftPoint | horzLine.p1 |
| Point | rightPoint | horzLine.p2 |
| Point | bottomPoint | vertLine.p2 |
| … | | |

| (d) Partial Shape Description for 古 | | |
|---|---|---|
| **components:** | | |
| Ten | ten | |
| Mouth | mouth | |
| **constraints:** | | |
| SameX | ten.bottomPoint | mouth.topPoint |
| LeftOf | ten.leftPoint | mouth.leftPoint |
| RightOf | ten.rightPoint | mouth.rightPoint |
| **aliases:** | | |
| … | | |

**Figure 3.** (a)(b) Visual structure and selected labels for the "ten" and "ancient" kanji. (c)(d) Corresponding partial LADDER shape descriptions for the "ten" and "ancient" kanji.

### Post-processing: Temporal Data and Aliasing

This section is dedicated to explaining how the Hashigo sketch interactive system can achieve the type of written technique recognition that is lacking in other kanji-based CAI systems. The post-processing step for written technique recognition encompasses stroke order and stroke direction critique, and it occurs subsequent to visual structure recognition. The idea behind this mechanism requires enumerating the strokes and points in the kanji shape descriptions explicitly, with labels indicating the target stroke order and direction. Figure 4 illustrates how the given alternative alias labels compare to Figure 3. First, this information is used during post-processing by referring back to the timing data of the lines and points segmented by the Sezgin recognizer. Since we already have stroke enumerated labels assigned from the constraints specification, we reference the timing data on those line and points by their labels. Lastly, we determine whether the enumerated stroke and point labels are listed in ascending temporal order. If there is a discrepancy in the ordering, this implies that the user had written the stroke or its direction differently from the target technique.

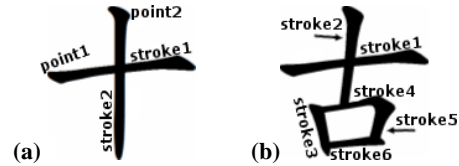

| (c) Partial Shape Description for 十 | | |
|---|---|---|
| **components:** | | |
| Line | horzLine | |
| Line | vertLine | |
| **constraints:** | | |
| … | | |
| **aliases:** | | |
| Line | stroke1 | horzLine |
| Line | stroke2 | vertLine |
| Point | point1 | horzLine.p1 |
| Point | point2 | vertLine.p2 |

**Figure 4.** (a)(b) Visual structure and selected labels for the "ten" and "ancient" kanji, respectively. (c) Corresponding partial LADDER shape description for the "ten" kanji.

### Application

We developed a fully operational learning tool which incorporates the recognition techniques described in the previous section. This application follows the teaching methods established in (Heisig 2007) by prompting users to sketch the kanji and elements (i.e., the corresponding simpler kanji contained within each kanji) introduced in the chapters. In our current system, we have developed a review setup for three of the chapters in the textbook, but can easily include additional chapters in the next version of our learning tool.

Upon initial usage of our application, users can choose whether to practice writing kanji in a particular chapter, or writing elements introduced in earlier chapters that are contained in the current chapter's kanji. After a selection is made, the user is shown an element or kanji and prompted to write it. Throughout the process, the user is given real-time feedback on their writing. If the application recognizes the visual structure of the kanji, the input will highlight the strokes to indicate visual correctness.

After the user is satisfied with the visual structure critique, the user runs the application for kanji, stroke order, stroke direction, and element sequence correctness. After written technique correctness has been processed, the appropriate feedback window will appear to provide valuable information to the user (Figure 5). Finally, at the end of the review, the user is given a final results window summarizing the user's performance (Figure 6).

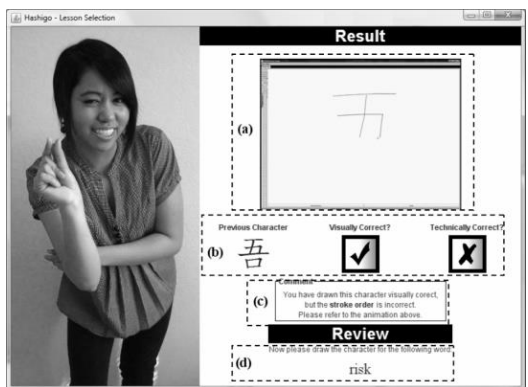

**Figure 5.** Feedback window with highlighted panels when user is reviewing kanji: (a) response panel, (b) critique panel, (c) comment panel, and (d) panel showing next kanji to sketch.

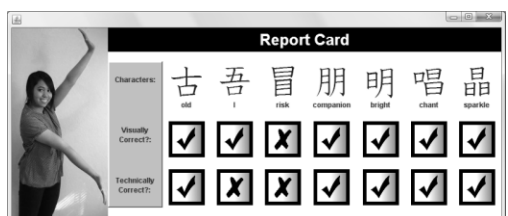

**Figure 6.** Feedback window showing a report card of sketch kanji in a particular lesson.

## Evaluation

We first evaluated our system on its visual structure recognition capabilities. The user study comprised of eleven international graduate students whom were proficient in kanji writing each of the nineteen kanji from a particular chapter twice. Since we wish to have model kanji for teaching students the correct way to write, our only requirement given to our participants was to write the kanji as if though they were teaching someone not familiar with them. The result of the user study was that our system correctly classified 92.9% of the provided kanji. The entire data from the user study was later used to tweak our shape descriptions in order to reflect natural human handwriting of kanji. Accuracies from existing vision- and gesture-based online kanji recognizers listed in (Liu *et al.* 2004) ranged from 85% to above 95%; so, the accuracy generated from our system is comparable when recognizing expert kanji users' handwriting.

The second evaluation focused on the written technique capabilities of our system, which involved determining whether our system could properly differentiate written technique factors like a human instructor could. The corresponding user study consisted of five non-East Asian students from the graduate school with no prior knowledge of kanji writing. We ran an initial user study on our participants by asking them to write seven prompted kanji from a given kanji lesson, giving them no further instruction than the visual structure of the kanji. When we provided this initial data to our system, our system generated 98.6% accuracy on the visual structure. We attribute this rise in accuracy to the higher care novice participants took in drawing the kanji exactly as presented, in contrast to their expert counterparts, who may have taken less care and who's previous writing habits may have biased their visual structure. In terms of written technique recognition, we first note that all the novice participants only gave correct written technique for 5.7% of the visually correct recognized kanji, which solidified the necessity of a sketch-based CAI system for teaching correct written technique. Secondly, our system perfectly differentiated those kanji with correct written technique from incorrect ones; that is, our system achieved 100% accuracy for written technique correctness.

Lastly, we evaluate the viability of a learning tool incorporating our system. The same novice users were asked to use our learning tool three times (i.e., preview, learn, and review) for a given lesson. After their third use of the learning tool, we conducted a final user study by collecting writing samples of the participants to gauge their kanji comprehension performance. After running through this last set of data, the novice users scored 100% accuracy on visual correctness and 97.1% on written technique correctness. This is a significant improvement of 5.7% of written technique correctness by these same users prior to using our learning tool.

## Future Work

Our system achieved reasonable recognition rates based on the data supplied from the user studies, but we plan on conducting additional user studies to tweak both existing and future kanji shape descriptions as we expand on the content of our learning tool. Additional data will allow us to make the system more robust to correctly classify messier kanji writing. Since our learning tool is for pedagogical purposes, we also plan to construct shape descriptions that restrict the level of messiness of written kanji, in order to discourage novice JSL students from neglecting correct visual structure form at the early kanji learning stage. In addition, we plan to eventually upgrade our recognition system with the upcoming new version of LADDER, which contains additional features that can improve the capabilities of our visual structure and written technique recognition methods.

Aspects of our learning tool were implemented based on comments given by instructors in the East Asian language program at our university. The program has recently planned on purchasing Wacom Cintiq monitors for classroom lab use so that students can use our learning tool in an optimal environment. We hope to continue working with members of the language program to mold the learning tool that better suits the needs of the JSL instructors. Additionally, we hope to showcase our software in a conference demonstration session to teach people Japanese kanji with our application.

# Conclusion

Language is a fundamentally human activity, as it is a tool for humans to express themselves. Therefore, there is no better way to instruct students of the Japanese language than through human Japanese language instructors. While no machine can hope to replace the value provided by a human instructor anytime soon, there is definitely room for CAI systems to augment JSL instruction and streamline the trivial yet time-consuming aspects. With our Hashigo sketch interactive learning tool for kanji instruction, our system can automate the type of instructional aid that human instructors already provide. Students will not only have a mechanism to improve their kanji comprehension outside the classroom through sketch interaction, but will also receive valuable feedback and assessments of their kanji writing to eliminate the early bad learning habits that reduce the effectiveness of learning kanji.

# Acknowledgments

This research is supported by the NSF IIS Creative IT Grant #0757557 Pilot: Let Your Notes Come Alive: The SkRUI Classroom Sketchbook.

The authors would like to thank members of the Sketch Recognition Laboratory, as well as George Adams, Mark Fransciso, Chiaki Johnson, Kazue Kurokawa, Ayaka Ono, and Krista Thom for their valuable assistance in this research.